\documentclass[aps,prl,showpacs,twocolumn,english,superscriptaddress,floatfix]{revtex4}
\usepackage[T1]{fontenc}
\usepackage[utf8]{luainputenc}
\usepackage{amssymb}
\usepackage{graphicx}
\usepackage{babel}
\usepackage{wasysym}
\usepackage{color}

\begin{document}

\title{Highly Retrievable Spinwave-Photon Entanglement Source}

\author{Sheng-Jun Yang}
\affiliation{Hefei National Laboratory for Physical Sciences at Microscale and Department of Modern Physics, University of Science and Technology of China, Hefei, Anhui 230026, China}
\affiliation{CAS Center for Excellence and Synergetic Innovation Center in Quantum Information and Quantum Physics, University of Science and Technology of China, Hefei, Anhui 230026, China}

\author{Xu-Jie Wang}
\affiliation{Hefei National Laboratory for Physical Sciences at Microscale and Department of Modern Physics, University of Science and Technology of China, Hefei, Anhui 230026, China}
\affiliation{CAS Center for Excellence and Synergetic Innovation Center in Quantum Information and Quantum Physics, University of Science and Technology of China, Hefei, Anhui 230026, China}

\author{Jun Li}
\affiliation{Hefei National Laboratory for Physical Sciences at Microscale and Department of Modern Physics, University of Science and Technology of China, Hefei, Anhui 230026, China}
\affiliation{CAS Center for Excellence and Synergetic Innovation Center in Quantum Information and Quantum Physics, University of Science and Technology of China, Hefei, Anhui 230026, China}

\author{Jun Rui}
\affiliation{Hefei National Laboratory for Physical Sciences at Microscale and Department of Modern Physics, University of Science and Technology of China, Hefei, Anhui 230026, China}
\affiliation{CAS Center for Excellence and Synergetic Innovation Center in Quantum Information and Quantum Physics, University of Science and Technology of China, Hefei, Anhui 230026, China}

\author{Xiao-Hui Bao}
\affiliation{Hefei National Laboratory for Physical Sciences at Microscale and Department of Modern Physics, University of Science and Technology of China, Hefei, Anhui 230026, China}
\affiliation{CAS Center for Excellence and Synergetic Innovation Center in Quantum Information and Quantum Physics, University of Science and Technology of China, Hefei, Anhui 230026, China}

\author{Jian-Wei Pan}
\affiliation{Hefei National Laboratory for Physical Sciences at Microscale and Department of Modern Physics, University of Science and Technology of China, Hefei, Anhui 230026, China}
\affiliation{CAS Center for Excellence and Synergetic Innovation Center in Quantum Information and Quantum Physics, University of Science and Technology of China, Hefei, Anhui 230026, China}

\begin{abstract}
Entanglement between a single photon and a quantum memory forms the building blocks for quantum repeater and quantum network. Previous entanglement sources are typically with low retrieval efficiency, which limits future larger-scale applications. Here, we report a source of highly retrievable spinwave-photon entanglement. Polarization entanglement is created through interaction of a single photon with ensemble of atoms inside a low-finesse ring cavity. The cavity is engineered to be resonant for dual spinwave modes, which thus enables efficient retrieval of the spinwave qubit. An intrinsic retrieval efficiency up to 76(4)\% has been observed. Such a highly retrievable atom-photon entanglement source will be very useful in future larger-scale quantum repeater and quantum network applications.
\end{abstract}

\pacs{03.67.Bg, 42.50.Dv, 03.67.Hk}

\maketitle

Quantum memories are of crucial importance for quantum repeater~\cite{Briegel1998}, quantum network~\cite{Kimble2008} and linear optical quantum computing~\cite{Kok2007}. Physical realization of quantum memory~\cite{Simon2010} can be either a single atom or an ensemble of single atoms. In ensemble-based quantum memories~\cite{Sangouard2011, Tittel2010}, the collective interference of a large number of atoms enables strong coupling with single photons efficiently~\cite{Duan2001, Fleischhauer2005}, which is a key advantage over the single atom counterparts. An entanglement pair between a photonic qubit and an ensemble based memory qubit forms the building blocks for quantum repeater and quantum network. Such atom-photon entanglement can be either generated directly through atom-photon interaction~\cite{Matsukevich2005, Chen2007, Dudin2010} or storage of photonic entanglement in an ensemble based quantum memory~\cite{Clausen2011, Saglamyurek2011, Zhang2011}. With atom-photon entanglement, significant progress has been made in realizing some functional elements of quantum network and quantum repeater, e.g., quantum teleportation from light to a matter~\cite{Chen2007,Bussieres2014a}, from matter to matter~\cite{Bao2012pnas}, and the realization of a quantum repeater note~\cite{Chou2007, Yuan2008}. However, the realization of quantum network with larger scale is still very challenging, e.g., entangling more than two remote quantum memories, or realizing entanglement purification in the quantum repeater architecture. Low retrieval efficiency for the atom-photon entanglement is a major limiting issue. On the other hand, progress towards efficient quantum memory with atomic ensembles has also been made significantly either with large atomic optical depth~\cite{Hedges2010, Hosseini2011, Chen2013} or using cavity enhancement~\cite{Simon2007, Bao2012cavity, Sabooni2013, Jobez2014a}. Nevertheless, a pair of atom-photon entanglement with high retrieval efficiency (>50\%) has not been realized so far.

In this paper, we report a source of spinwave-photon entanglement with a high retrieval efficiency. Polarization entanglement is created through interaction of a single photon with ensemble of atoms inside a low-finesse ring cavity. We carefully tune the polarization birefringence inside the cavity to realize simultaneous resonance for dual spinwave modes. The intrinsic retrieval efficiency---also called as ``peak single-quantum conversion efficiency'' in Ref.~\cite{Simon2007} and ``conditional retrieval efficiency'' in Ref.~\cite{Kimble2008}---from a stored spinwave qubit to a photonic qubit is enhanced up to $76(4)\%$. Such a high efficiency significantly exceeds the free-space single-photon results~\cite{Laurat2006,Radnaev2010}, and is comparable to the previous results with cavity enhancement~\cite{Simon2007,Bao2012cavity}. The spinwave coherence lifetime is measured to be $25.2(4)\,\mu\rm{s}$, which can be extended significantly by using optical lattice confinement~\cite{Radnaev2010,Dudin2010}. In comparison with atom-photon entanglement in the single-atom based quantum memories~\cite{Weber2009, Stute2012, Gao2012, DeGreve2012}, our system features the ease of preparation and robust interaction with single-photons. We think the demonstrated highly retrievable spinwave-photon entanglement will have plenty of applications, such as remote entanglement of several quantum memories and scalable generation of large cluster-states~\cite{Barrett2010}.

\begin{figure}[h]
\includegraphics[width=\columnwidth]{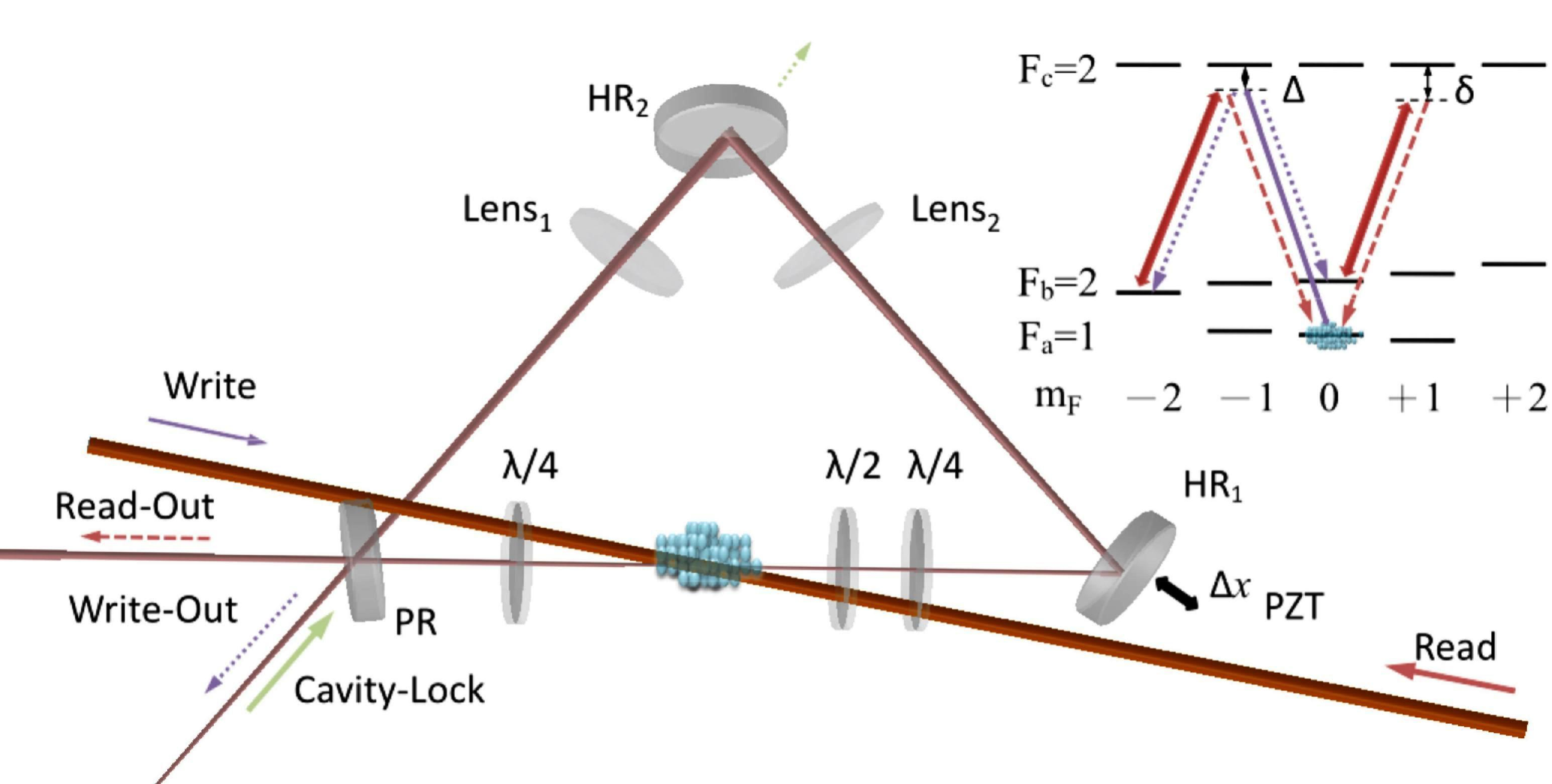}
\caption{(Color online) Schematic of the experimental setup. The counter-propagating write and read pulses intersect with the cavity mode through cold atomic ensemble with an angle of $\sim\!2.5^{\circ}$. All beams are in the same horizontal plane. For the write and read modes, beam waist is $\sim\!300\,\rm{\mu m}$. While for the write-out and read-out modes, beam waist is $\sim\!90\,\rm{\mu m}$. The ring cavity mainly consists of two planoconvex lens ($f\!=\!250\,\rm{mm}$), two highly reflecting mirrors $\rm{HR}_{1,2}$  ($R\!\geq\!99.9\%$) and one partially reflecting mirror PR ($R\!\simeq\!80\%$) as the write- and read-out signal photon coupling ports. The cavity-locking beam is combined along the write-out channel with a nonpolarized beam splitter (partial reflection $\sim\!95\%$). The quarter- and half-wave plates ($\lambda/4$ and $\lambda/2$) inside the cavity are used for polarization compensation; thus the write- and read-out photons and the cavity-locking pulse are all resonance with the cavity for arbitrary polarization. Leaks of the cavity-locking pulse from $\rm{HR}_{2}$ are detected with a homemade fast photodiode to generate the error signal, and feedback drive the mirror $\rm{HR}_{1}$ with a piezoelectric transducer (PZT) for cavity stabilization. Inset: energy levels of the $^{87}\!\rm{Rb}$ atoms. The atoms are initially prepared in ground state $\left|F\!=\!1,m_{F}\!=\!0\right\rangle $. The write and read beams are both red detuned by 40\,MHz relative to the atomic transitions . The two sub-levels $\left|2,-2\right\rangle $ and $\left|2,0\right\rangle $ are used for coherence storage of atomic spinwave excitations.}
\label{fig:Setup}
\end{figure}

The experimental setup is shown in Fig.\,\ref{fig:Setup}. We prepare an ensemble of $^{87}\!\rm{Rb}$ atoms through $30\,\rm{ms}$ magneto-optical trapping, which gives a temperature of $\sim\!20\,\rm{\mu K}$. In the subsequent $2\,\rm{ms}$ we repeat the entanglement creation and verification trials. In the beginning of each trial, atoms are optically pumped to the initial state of $5S_{1/2}\!:\!\left|F\!=\!1,m_{F}\!=\!0\right\rangle$.
To generate a pair of spinwave-photon entanglement, we make use of the spontaneous Raman scattering process~\cite{Duan2001}. The energy levels employed are shown as an inset of Fig.\,\ref{fig:Setup}. When applying a $\sigma^-$-polarized write beam, the interference between two scattering channels gives rise to entanglement between the polarization of the scattered write-out photon and the internal state of the atomic spinwave excitation in the form of
\begin{equation}
\left|\Psi\left(t\right)\right\rangle\!=\!\sin\eta\left|\rm{L}\right\rangle S_{0,0}^{\dagger}\!-\!\cos\eta\, \rm{e}^{i\varphi\left(t\right)}\left|\rm{R}\right\rangle \emph{S}_{0,-2}^{\dagger},
\end{equation}
where $\sin\eta\!=\!\sqrt{3/5}$ is the relevant Clebsch-Gorden coefficients of the energy levels, $\left\{\rm{R/L}\!:=\!\sigma^{\pm}\right\}$ is circular polarization, $S_{i,j}^{\dagger}$ is the collective atomic spinwave excitation between the ground states $\left|1,i\right\rangle $ and $\left|2,j\right\rangle $, and phase oscillation $\varphi(t)\!=\!-2\mu_{B}Bt$ is induced by the bias magnetic field $B$ ($\sim\!0.6\,\rm{Gauss}$). We set the write pulse power of $\sim\!4\,\rm{\mu W}$ and pulse width (FWHM) of $80\,\rm{ns}$, which result in a detected write-out probability about $7.7\permil$. After a storage time $t$, we convert the stored spinwave excitation back to a single photon by applying a strong $\sigma^+$-polarized read pulse with a power of $\sim\!300\,\rm{\mu W}$ and the FWHM of the read-out signal is about $80\,\rm{ns}$. Afterwards, regular polarization measurement is made on the read-out photon to verify the spinwave-photon entanglement.

In order to achieve a high spinwave-to-photon conversion efficiency, we make use of a ring cavity. The cavity's fuction is two-fold. First, it enhances the process of writing and enables the creation of a spinwave excitation with perfect single-mode quality. Second, the cavity enhances the reading process through Purcell effect and improves the retrieval efficiency significantly for small ensembles~\cite{Gorshkov2007, TanjiSuzuki2011}. The angle separation between the cavity mode and the write/read direction is $2.5^{\circ}$. The write- and read-out photons are counter-propagating, satisfying the phase matching condition $\mathbf{k}_{\rm{w}}\!+\!\mathbf{k}_{\rm{r}}\!=\!\mathbf{k}_{\rm{wo}}\!+\!\mathbf{k}_{\rm{ro}}$ and resonating with the cavity. All the optical components of the cavity are outside the vacuum glass cell, and surface-coated to reduce intra-cavity losses. The vacuum glass cell is also anti-reflecting coated both sides. The coupling mirror (PR) has a reflection rate of $80\%$ for both horizontal and vertical polarizations. The measured free spectral range (FSR) is about $489.6\,\rm{MHz}$ and the finesse is $\mathcal{F}\!\sim\!18$. The cavity-locking pulse, which is one FSR blue-detuned with the $D1$-line $\left|F_{b}\!=\!2\right\rangle\!\leftrightarrow\!\left|F_{c}\!=\!2\right\rangle$, is backwards combined along the write-out channel with a $95\%$ partial-reflecting nonpolarized beam combiner to actively stabilize cavity frequency resonance via the Pound-Drever-Hall method. As there is some birefringence due to three dielectric mirrors and the glass cell inside the cavity, orthogonal linearly polarizations will not be resonate with the ring cavity simultaneously for a given frequency. Thus we insert two quarter-wave plates and one half-wave plate inside the cavity for phase compensation. After fine adjustment of the wave plates, arbitrary polarizations are resonant with the cavity simultaneously. The write- and read-out photons are collected by single-mode optical fibers, subsequently filtered in frequency, and finally detected with four avalanche single-photon detectors.

\begin{figure}[h]
\includegraphics[width=\columnwidth]{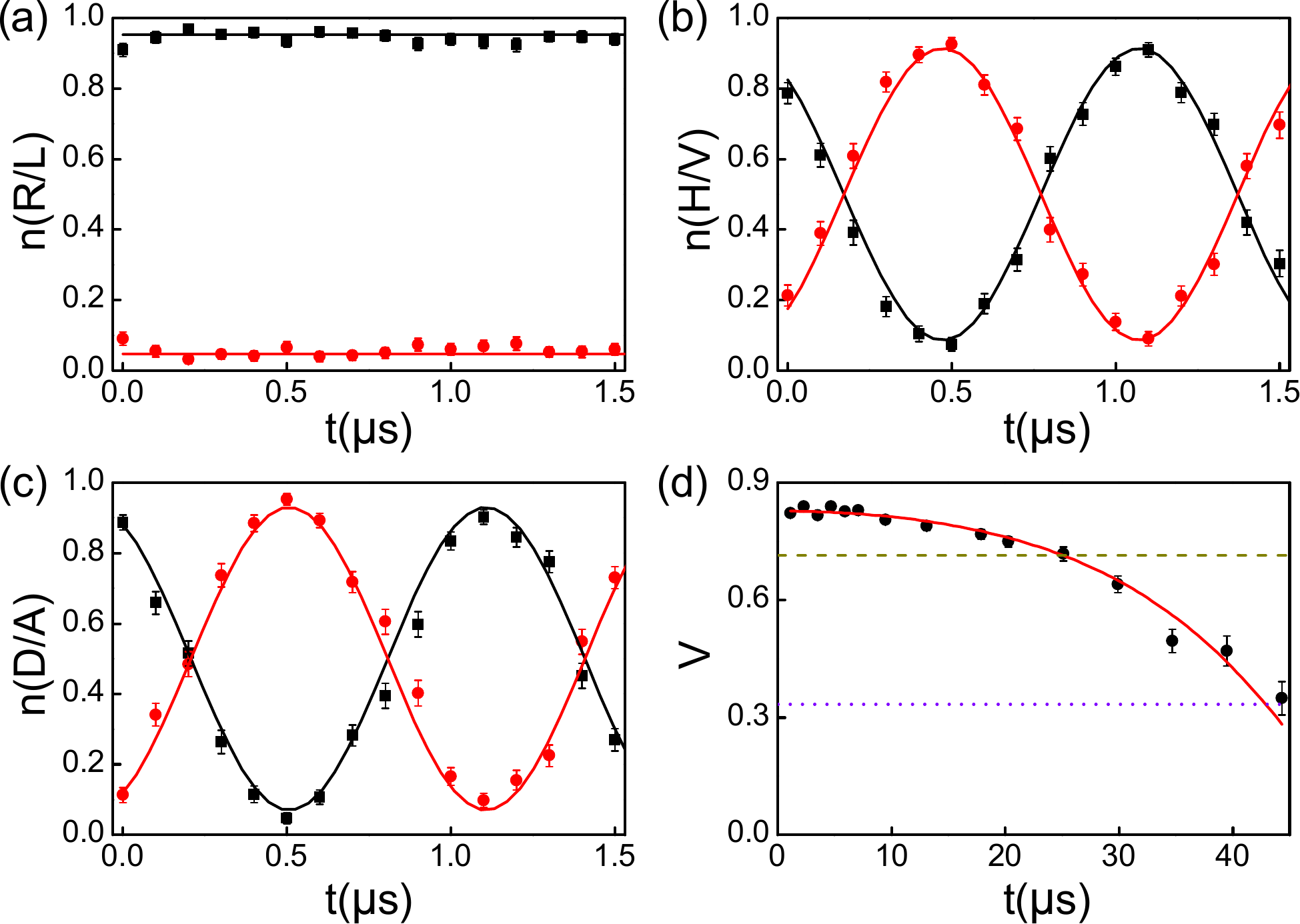}
\caption{(Color online) Measurement of polarization correlations. (a)-(c) The normalized coincidence counts of the write- and read-out photons for the three polarization settings $\left\{\rm{R/L,H/V,D/A}\right\}$ during the initial $1.5\,\rm{\mu s}$ storage, where the coincidence count rate is  about $4.3\!\times\!10^{-4}$ per pulse. Results for the setting of $\rm{R/L}$ are linearly fitted, while results for the setting of $\rm{H/V}$ and $\rm{D/A}$ are fitted by a damped sinusoidal function with an oscillation period of $1.2\,\rm{\mu s}$. (d) The averaged polarization visibility for different storage time, with a fitting result of initial visibility $V(t\!=\!0\,\rm{\mu s})\!\simeq\!0.83(0)$ and decay lifetime $\tau\!\simeq\!33.2(8)\,\rm{\mu s}$. The yellow (dashed) line corresponds to $V=0.71$, and the violet (dotted) line corresponds to $V=0.33$.}
\label{fig:Visibility}
\end{figure}

We first measure the polarization correlations between the write- and read-out photons to characterize the photon-spinwave entanglement.
By applying the read beam, the photon-spinwave entanglement is converted a pair of phonic entanglement between the write-out photon and read-out photons in the form of $\left|\Psi\left(t\right)\right\rangle'\!=\!\sin{\eta}\left|\rm{LR}\right\rangle\!-\!\cos{\eta} \,\rm{e}^{i\varphi\left(t\right)}\left|\rm{RL}\right\rangle$.
Normalized coincidence counts in three different polarization settings $\left\{\rm{R/L},\rm{H/V}\!:=\!0^{\circ}/90^{\circ},\rm{D/A}\!:=\!\pm45^{\circ}\right\}$ are shown in Fig.\,\ref{fig:Visibility}\,(a)\,-\,(c) as a function of storage time $t$, where the black square data
refers to the coincidence count for perpendicular polarizations $n_\perp\!=\!n_{\rm{RL(HV,DA)}}\!+\!n_{\rm{LR(VH,AD)}}$ and the red circle data refers to the coincidence count for parallel polarizations $n_\parallel\!=\!n_{\rm{RR(HH,DD)}}\!+\!n_{\rm{LL(VV,AA)}}$. Measurement result in the $|R/L\rangle$ setting identifies the two major components of $|LR\rangle$ and $|RL\rangle$, while measurements in the superposition settings of $|H/V\rangle$ and $|D/A\rangle$ identify the time-dependent phase between $|LR\rangle$ and $|RL\rangle$ due to Larmor precession. Defining the polarization visibility as
\begin{equation}
V_{\{\rm{RL},\rm{HV},\rm{DA}\}}\!=\!\left|n_\perp\!-\!n_\parallel\right|/\left(n_\perp\!+\!n_\parallel\right),
\end{equation}
we obtain the fitting visibilities for the three polarization settings as $\{0.906(4),0.830(11),0.860(15)\}$, where the value of the $\rm{H/V}$ and $\rm{D/A}$ settings corresponds to the maximum oscillated points. These visibilities fulfill the condition of $V\geq0.71$ to violate the Bell-CHSH inequality. As noticed, the visibility at $t=0$~$\rm{\mu s}$ is not maximized, which is due to an initial peak separation of $\sim\!110\,\rm{ns}$ between the write- and read-out pulses. Compared with the theoretical upbound visibility of $V\!\simeq\!0.98$, the measured visibilities are mainly limited by background noise of the write- and read-out channels, with averaged detection rates of $0.6\permil$ and $3.9\permil$ respectively, which could be further improved by, e.g., enlarging the angle between the write/read beam and the cavity mode. For an acceptable low noise level, smaller write-out photon emanation rate and better measured retrieval efficiency would further obtain a larger cross correlation $g_2$, and thus higher fidelity of entanglement~\cite{Chen2007}. We also measured the storage coherence time of the average visibility $V\!=\!(V_{\rm{RL}}\!+\!V_{\rm{HV}}\!+\!V_{\rm{DA}})/3$ in Fig.\,\ref{fig:Visibility}\,(d), where the selective measurement points correspond to the maximum visibilities of all the three polarization settings. Fitting with the function of $V(t)\!=\!1\!-\!2\!/\!(ae^{-\!t^{2}\!/\!\tau^{2}}\!+\!1)$, we got the results of initial visibility $V(t\!=\!0\,\rm{\mu s})\!\simeq\!0.83(0)$ and lifetime $\tau\!\simeq\!33.2(8)\,\rm{\mu s}$.

\begin{table}[h]
\protect\caption{Measurement of the CHSH $S$ parameter.}
\centering
\begin{tabular*}{\columnwidth}{@{\extracolsep\fill}lccc}
\toprule
$t_{\rm{store}}(\rm{\mu s})$ & 1.1 & 9.5 & 17.9\\
\hline
S-value & 2.30(3) & 2.20(3) & 2.11(4)\\
S.D. & 10.4 & 5.8 & 2.6\\
\toprule%
\end{tabular*}
\label{tab:CHSH}
\end{table}

Next, we measure the Bell-CHSH inequality to confirm the photon-spinwave entanglement. The $S$ parameter is defined as
\begin{equation}
S\!=\!\left|E(\theta_{1},\theta_{2})\!-\!E(\theta_{1},\theta'_{2})\!-\!E(\theta'_{1},\theta_{2})\!-\!E(\theta'_{1},\theta'_{2})\right| \end{equation}
with correlation function $E(\theta_{i},\theta_{j})$ given by
\begin{equation}
\frac{n(\theta_{i},\theta_{j})\!+\!n(\theta_{i}^{\bot},\theta_{j}^{\bot})\!-\!n(\theta_{i}^{\bot},\theta_{j})\!-\!n(\theta_{i},\theta_{j}^{\bot})}{n(\theta_{i},\theta_{j})\!+\!n(\theta_{i}^{\bot},\theta_{j}^{\bot})\!+\!n(\theta_{i}^{\bot},\theta_{j})\!+\!n(\theta_{i},\theta_{j}^{\bot})},
\end{equation}
where $n(\theta_{i},\theta_{j})$ is the coincidence counts and $\theta^{\bot}\!=\!\theta\!+\!\frac{\pi}{2}$.
For any local realistic theory, the $S$ parameter can not be larger than 2. A polarization measurement setting of $\{\theta_{1},\theta'_{1},\theta_{2},\theta'_{2}\}\!:=\!\{0^{\circ},45^{\circ},22.5^{\circ},-22.5^{\circ}\}$ corresponds to the maximum violation for Bell states. The maximum value for the $S$ parameter is $\left(1\!+\!\sin2\eta\right)\sqrt{2}\!\sim\!2.8$. Due to the Larmor precession, the measured points of the storage time in Table\,\,\ref{tab:CHSH} are selected for maximum visibility of all the three settings $\left\{\rm{R/L},\rm{H/V},\rm{D/A}\right\}$ as shown in Fig.\,\ref{fig:Visibility}, that most close to the Bell state $\left|\Psi^{\pm}\right\rangle $. At the beginning of the storage, the measured $S$ parameter is $2.30(3)$ which violates the inequality $S\!\leq\!2$ by ten standard deviations, which confirms the entanglement between the atomic spinwave and the photonic qubit. At the storage time of $18\,\mu s$, we still get $S\!=\!2.11(4)\nleq2$ with $2.6$ standard deviations.

\begin{figure}[h]
\includegraphics[width=.8\columnwidth]{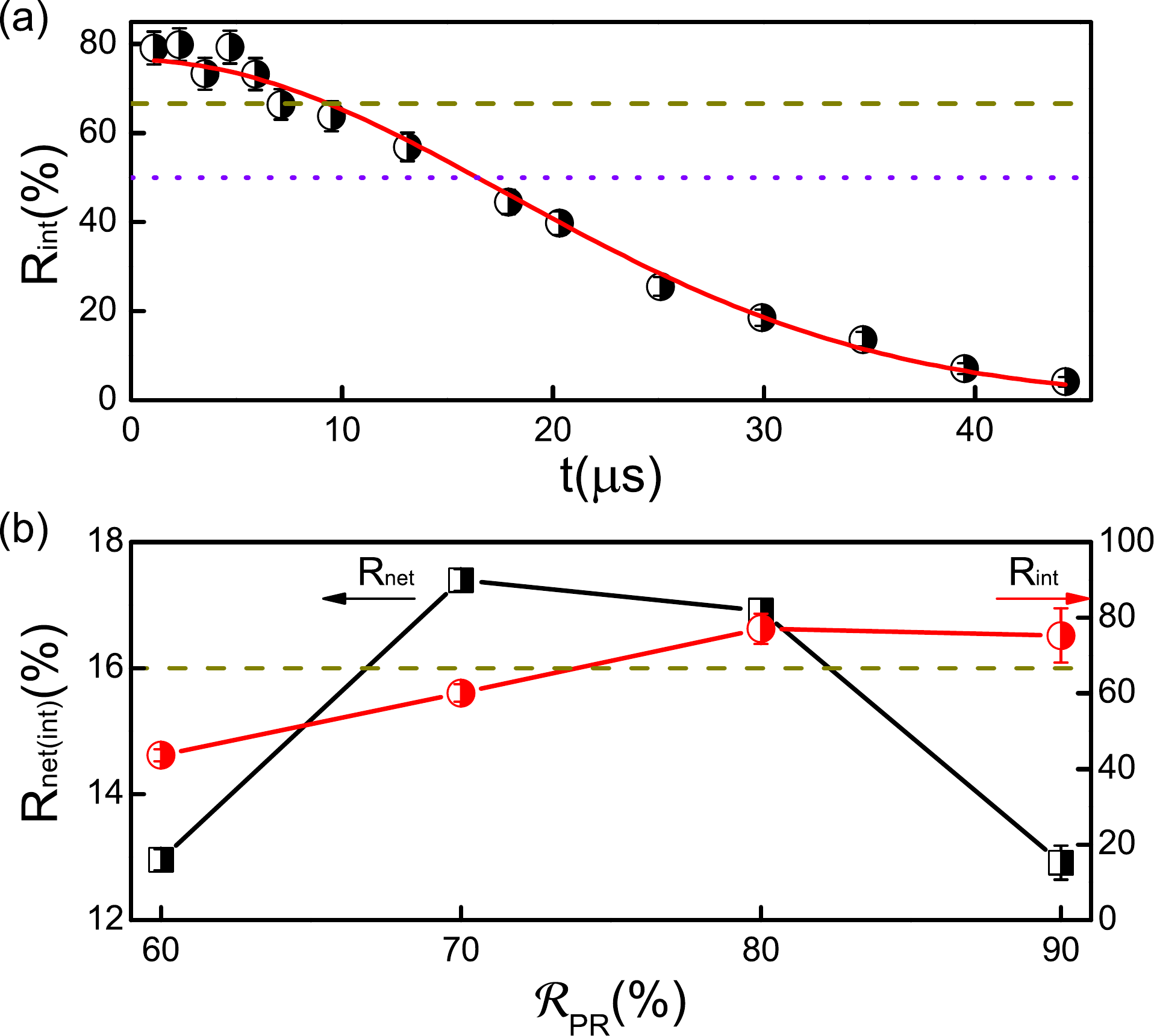}
\caption{(Color online) (a) Measured time dependence of the intrinsic retrieval efficiency $R_{\rm{int}}$. The yellow (dashed) line corresponds to an efficiency of $66.7\%$, and the violet (dotted) line corresponds to an efficiency of $50\%$. (b) Measured retrieval efficiencies as a function of cavity mirror reflection rate $\mathcal{R}_{\rm{PR}}$. Data points in black square correspond to the net retrieval efficiency $R_{\rm{net}}$, while data points in red circle correspond to the intrinsic retrieval efficiency $R_{\rm{int}}$.}
\label{fig:Efficiency}
\end{figure}

Finally we measure the spinwave-to-photon conversion efficiency. In our experiment, we first measure the net retrieval efficiency $R_{\rm{net}}\!=\!(p_{\rm{coin}}\!-\!p_{\rm{wo}}p_{\rm{ro}})/(p_{\rm{wo}}\!-\!p_{\rm{wobg}})$, where $p_{\rm{wo},\rm{ro}}$ is the measure write- and read-out probability, $p_{\rm{coin}}$ is the coincidence rate between them, $p_{\rm{wobg}}$ is the background noise in the write-out channel without atoms, and then calculate the intrinsic retrieval efficiency $R_{\rm{int}}$ inside the cavity. For $\mathcal{R}_{\rm{PR}}\!\simeq\!80\%$, the initial measured retrieval efficiency $R_{\rm{net}}\left(t\!=\!0\,\rm{\mu s}\right)$ is about $16.9(2)\%$, corresponding to an intrinsic retrieval efficiency $R_{\rm{int}}\!\sim\!76(4)\%$, after correcting all kind of losses of the read-out photon (mainly including cavity losses $\mathcal{L}\!\sim\!11(1)\%$, transmittance of the read-out channel $\mathcal{T}\!\sim\!65(3)\%$, and detection efficiency of the avalanche photodiode $\mathcal{D}\!\sim\!63(2)\%$). Comparing with the typical value of $20\%$ in our experiment without using a cavity, the intrinsic retrieval efficiency gets increased by 3.8 times. We also measure the time dependence of retrieval efficiency, with the result shown in Fig.~\ref{fig:Efficiency}(a). $1/e$ lifetime is measured to be $25.2(4)\,\rm{\mu s}$, and $R_{\rm{int}}\left(9.3\,\rm{\mu s}\right)\!\simeq\!66.7\%$, $R_{\rm{int}}\left(16.4\,\rm{\mu s}\right)\!\simeq\!50\%$. By replacing the coupling cavity mirror PR with different reflections ($60\!-\!90\%$), we also show the retrieval efficiency of $R_{\rm{net}}$ and $R_{\rm{int}}$ for different cavity finesses ($\mathcal{F}\!\sim\!10\!-\!30$) in Fig.\,\ref{fig:Efficiency}\,(b). The results are averaged for the two circular polarization settings $\left|\rm{LR}\right\rangle$ and $\left|\rm{RL}\right\rangle$. The black square data is the measured net retrieval efficiency $R_{\rm{net}}$, while the red circle one is the intrinsic retrieval efficiency $R_{\rm{int}}$. For the PR-mirror reflection $80\%$ and $90\%$, $R_{\rm{int}}$ are about $76\%$, larger than the threshold of $66.7\%$ which is required to violate the Bell inequality without detection loophole~\cite{Eberhard1993}. The droop of the measured retrieval efficiency for high cavity finesse is due to the increasing of relative weight of intra-cavity loss over $\mathcal{R}_{\rm{PR}}$.

Of course, to achieve a high net retrieval efficiency $R_{\rm{net}}$, we need to make more efforts to reduce all kinds of losses for the read-out photons. By fine matching write(-out) and read(-out) modes, using more compact and stable cavity design, and also larger atomic optical depth, an intrinsic retrieval efficiency higher than $85\%$ will be reachable. Assuming photon losses $\le\!0.1\%$ for each optical surface within the cavity, the total intra-cavity loss is $\mathcal{L}\!\le\!1.6\%$, corresponding to a signal photon emanation rate of $93\%$ out from the cavity when $\mathcal{R}_{\rm{PR}}\!\simeq\!80\%$. And the read-out photon could be sent to a single-photon detection directly after proper free-space noise filtering. Together with high-efficient superconductor transition edge detectors~\cite{Ramelow2013}, an overall measured net retrieval efficiency of $\sim70\%$ is foreseeable in the near future. In our present experiment, the spinwave coherence time is merely $\sim\!25\,\rm{\mu s}$, which is mainly limited by atom thermal motion and inhomogeneous energy shift of the ground states. It can be further improved by four orders up to hundreds of ms by using an optical lattice trap~\cite{Radnaev2010} and selecting two pairs of magnetic-insensitive atomic transitions~\cite{Xu2013}.

In conclusion, we have successfully generated a highly retrievable spinwave-photon entanglement with an ensemble of atoms in a ring cavity. Such an efficient atom-photon entanglement source may have lots of applications in quantum repeater and quantum network, like entangling more atomic ensembles~\cite{Barrett2010}, connecting two quantum repeater nodes, performing entanglement purification~\cite{Pan2003}. This source can also be used to create large cluster states~\cite{Walther2007, Yao2012} for linear optical quantum computing~\cite{Raussendorf2001, Kok2007}. Besides, our source may also be useful to perform a loophole-free test of Bell inequalities~\cite{Brunner2014}.

\begin{acknowledgments}
This work was supported by the National Natural Science Foundation of China, National Fundamental Research Program of China (under Grant No. 2011CB921300), and the Chinese Academy of Sciences. X.-H.B. acknowledges support from the Youth Qianren Program.
\end{acknowledgments}

\bibliography{myref}

\end{document}